\documentclass[
 aip,apl,
 amsmath,amssymb,reprint
]{revtex4-1}

\usepackage{graphicx}
\usepackage{dcolumn}
\usepackage{bm}
\usepackage{subfig}

\begin{document}

\preprint{AIP/123-QED}

\title[Effect of nano-scale surface roughness on transverse energy spread from GaAs photocathodes]{Effect of nano-scale surface roughness on transverse energy spread from GaAs photocathodes}

\author{Siddharth Karkare}
\affiliation{ Physics Department, Cornell University.}
\author{Ivan Bazarov}
\affiliation{ Physics Department, Cornell University.}

\begin{abstract}
High quantum yield, low transverse energy spread and prompt response time make GaAs activated to negative electron affinity (NEA), an ideal candidate for a photocathode in high brightness photoinjectors. Even after decades of investigation, the exact mechanism of electron emission from GaAs is not well understood. We show that a nano‐scale surface roughness can affect the transverse electron spread from GaAs by nearly an order of magnitude and explain the seemingly controversial experimental results obtained so far. This model can also explain the measured dependence of transverse energy spread on the wavelength of incident light.
\end{abstract}

\maketitle

The need for a high brightness electron beam is well established\cite{NIMA}. GaAs activated to negative electron affinity via cesiation is a high quantum efficiency (QE) photocathode and can be effectively used for producing such beams\cite{NIMA,main_Ivan}. Properties of GaAs as a photocathode have been studied for decades\cite{spicer,narrow_cone,transverse_spread}. However, the mechanism of photoemission from these photocathodes is not well understood.

Most models follow the Spicer 3-step theory\cite{spicer}, and they all assume near full thermalization of electrons to the $\Gamma$ valley minimum when excited with near band-gap energy photons. The difference arises when one considers the effects on the electron going through the surface (band bending and activation regions). To explain the experimental data, one approach argues that the electrons undergo sufficient scattering at the surface so that the transverse energies of the emitted electrons are of the order of 25 meV (thermal energy at room temperature)\cite{main_Ivan,transverse_spread}. The other body of work, however, treats the emission process as a refraction of a Bloch wave at an ideal surface while largely ignoring scattering effects at the surface. It predicts the transverse energy of the electrons to be around 1 to 2 meV at room temperature\cite{narrow_cone,Pollard} and the electrons are emitted in a cone with an half angle of $15^{\circ}$, which is a result of the small effective mass of the electrons in the $\Gamma$ valley of GaAs.

Furthermore, the experimental measurements of the mean thermal energy (MTE) and thermal emittance are also inconsistent. Some groups report values of MTE close to the room temperature thermal energies of 25meV \cite{main_Ivan,transverse_spread}. While others report values of measured MTE near 2meV and the $15^{\circ}$ angular distribution as predicted by the second model\cite{narrow_cone}. Additionally, measurements  show that MTE depends strongly on the wavelength of light used for photoemission\cite{main_Ivan}. None of the existing models can quantitatively explain this dependence. MTE and normalized transverse rms emittance ($\epsilon_{nx}$) are related to the spot size of the laser ($\sigma_{x}$) by 
$
\epsilon_{nx}=\sigma_{x}\sqrt{\mathrm{MTE}/\left(m_{e}c^{2}\right)}
$
where $m_{e}c^{2}$ is the rest mass energy of a free electron.

In this paper, we attempt to resolve these discrepancies by considering the effects of nano-scale surface roughness of GaAs on the MTE. The surface roughness effect can explain measurement data\cite{main_Ivan} and the variation of the MTE with incident wavelength, as well as reconciles seemingly contradictory collection of data in the literature\cite{narrow_cone, transverse_spread}.

Typical bulk GaAs preparation procedures include surface cleaning of heavily p-doped GaAs using high temperature cleaning and/or H-cleaning. GaAs wafers we used underwent the same treatment as in \cite{main_Ivan}. To achieve good QE, the samples are typically heat treated to around $580^{\circ}\mathrm{C}$ for 1-3 hours. NEA condition is achieved via a well-known `yo-yo' procedure which employs alternating exposure to Cs and $\mathrm{NF_{3}}$.
The surface of atomically polished GaAs was studied before and after activation. The surface was imaged and the  roughness was measured using atomic force microscopy (AFM).  It was found that the typical roughness of the surface before activation was less than 0.5nm (rms) (see Fig.1(a)). 

\begin{figure}
\includegraphics[width=0.85\linewidth,bb=40bp 400bp 300bp 790bp]{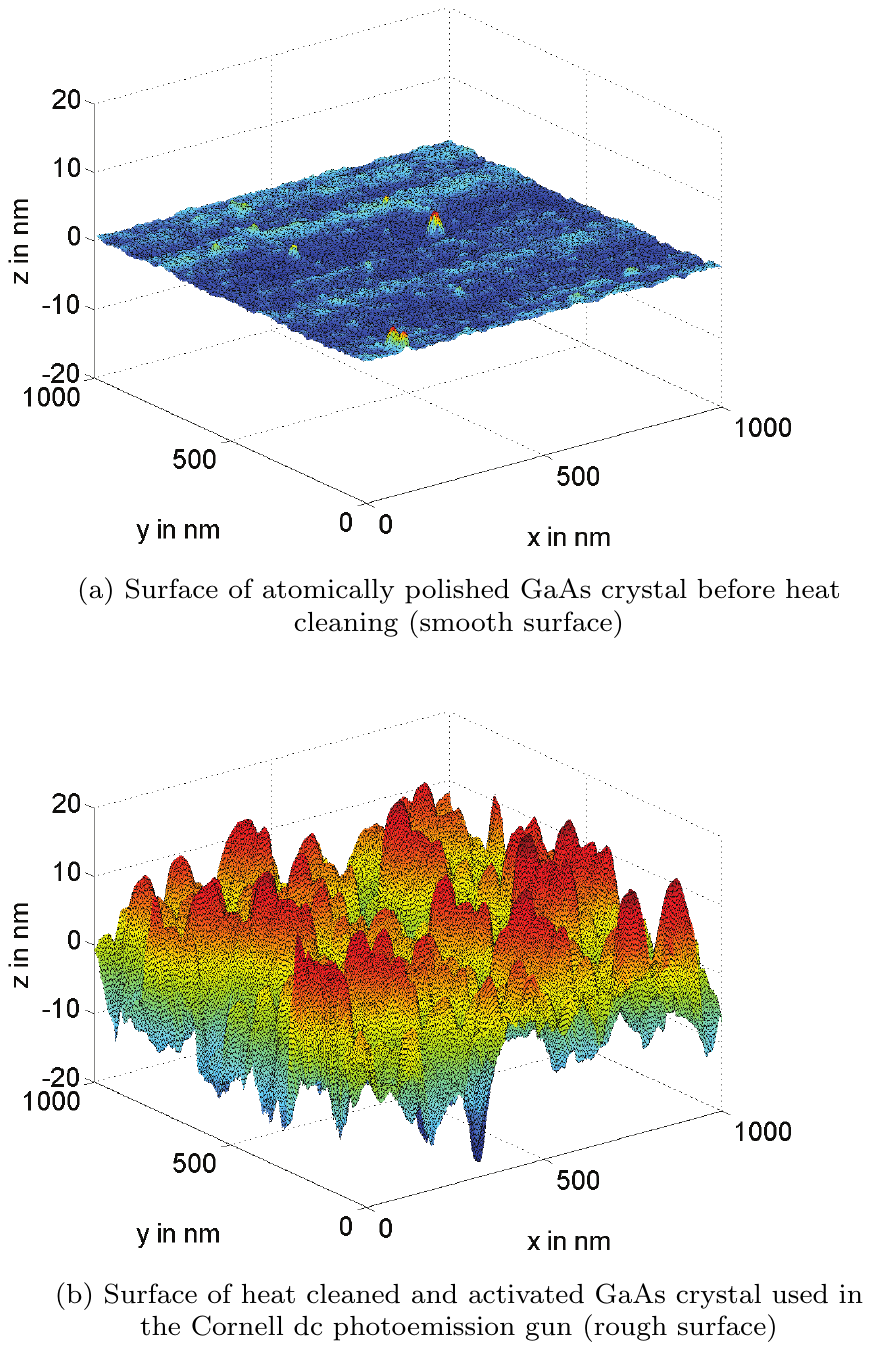}
\caption{AFM images of GaAs surfaces}
\end{figure}

The surface of an activated GaAs crystal, used in and removed from the Cornell DC photoemission gun\cite{bruce} was studied. The roughness of this surface was about 6nm (rms) (see Fig.1(b)). This roughness is typical of atomically polished GaAs surface that undergoes the usual heat-treatment procedure and is not detectable visually under an optical microscope. The mirror-like surface appearance of the surface is not affected at this roughness level. This nano-roughness develops as a result of thermal etching, surface faceting or the dissociation of GaAs which occurs at 580$^\circ$C.\cite{NEA_Semicon,dissociation}.

The effect of surface roughness on photoemission was previously examined\cite{Bradley}. The major effect of surface roughness is due to the electrons being emitted in a direction perpendicular to the local surface instead of the global normal to the surface. The transverse energy is a result of the transverse component of the velocity of the electrons being emitted perpendicular to the local surface. This is the slope effect\cite{Bradley}. The second effect is due to the bending of the electric field used to extract the electrons in the close vicinity of the rough surface. This is the field effect\cite{Bradley}. Assuming an extraction field of 3-5MV/m, as in the Cornell DC photoemission gun,  the electric field in the vicinity of the rough surface shown in Fig.1(b) was calculated\cite{Gorlov}.
 Ignoring the effects of scattering and assuming the valance bands to be flat, the energy of the emitted electrons can be written as
$
E'=\hbar\nu-E_{g}+E_{A}\label{eq:1}
$
where $\hbar\nu$ in the energy of the incident photons, $E_{g}=1.42\mathrm{eV}$ is the band gap in GaAs and $E_{A}$ is the negative electron affinity typically ranging from $0.1\mathrm{eV}$ to $0.25\mathrm{eV}$.  The higher is the energy of the emitted electron, the higher will be the longitudinal and transverse components of its velocity, implying a higher transverse energy. This explains the rise in MTE with the energy of incident photons.

\begin{figure}
\includegraphics[width=\linewidth,bb=50bp 185bp 655bp 562bp]{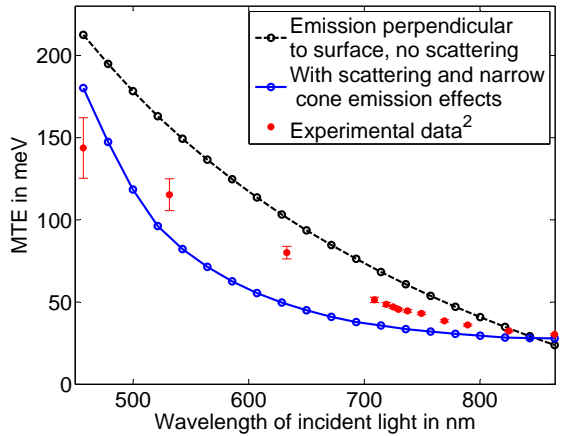}
\caption{\label{fig:3a}MTE vs wavelength of incident photons - experimental results and predictions of models taking into account surface roughness for surface shown in Fig.1(b)}
\end{figure}

In simulations, the electrons were launched from a $256\times256$ square grid on the surface shown in Fig.1(b), with the energy $E'$, in direction normal to the surface at the point of launch. The value of $E_{A}$ was chosen to be $0.145\mathrm{eV}$, to better match the experimental results.
Fig.\ref{fig:3a} shows the MTE as a function of the wavelength of incident light. The red points are the experimental data\cite{main_Ivan}. The dashed black curve is the curve obtained by launching electrons perpendicular to the rough surface shown in Fig.1(b) at a fixed energy $E'$. We see that this simple analysis produces a dependence of MTE on the wavelength that matches closely with the experimental values.

A more elaborate model has been developed, the result of which is shown by the solid line in Fig.\ref{fig:3a}. This model is described below. Due to energy and momentum conservation, electrons in the heavy-hole (hh), light-hole (lh) and split-off (so) valence bands absorb photons and are excited into the $\Gamma$ valley of the conduction band via vertical transitions. The hh and lh bands are assumed to be identical and parabolic\cite{GaAs_properties} with effective mass $m_{hh}=m_{lh}=0.45m_{e}$. The effective mass in the $\Gamma$ valley is $m_{\Gamma}=0.067m_{e}$. We ignore the excitations from the split-off band. From conservation of energy and assuming a vertical transition, the energy of the excited electron with respect to $\Gamma$ valley minimum is given by $E_{0}=\left(\hbar\nu-E_{g}\right)m_{hh}/(m_{hh}+m_{\Gamma})\label{eq:2}$. These electrons, excited into the $\Gamma$ valley diffuse towards the surface\cite{spicer1}. During this transport towards the surface they scatter with phonons. Only the polar phonons (optical and acoustic) have a significant effect\cite{tomizawa}. The excited electrons tend to lose energy by scattering with the polar optical phonons. The polar acoustic phonons are low energy phonons and do not cause a significant energy loss, but do give rise to the spread in the initial delta-function like energy distribution and cause the electrons to thermalize. The scattering rates for the polar optical phonon (35meV energy) are given in\cite{tomizawa}. The effect of polar acoustic phonons was modeled by considering a low energy phonon (1meV energy) with scattering cross-section same as that of the polar optical phonon\cite{Ridley}. These scattering rates were used to numerically calculate the time evolution of the electron energy distribution function, using a Monte-Carlo (MC) simulation.
For incident photon energy $\hbar\nu$, the number of electrons with energy $E$, at time $t$ after excitation is given by $f(E,\hbar\nu,t)$. In the above analysis, inter-valley and impurity scattering has been ignored. 
We assume that the time for excitation and emission from the surface is negligible compared to the time required for the transport to the surface. The number of electrons reaching the surface between time $t$ and $t+dt$ is
$
\eta\left(t,\hbar\nu\right)\cdot dt=(\delta P\left(t,\tau\right))/(\delta t)\cdot dt
$
,where $P\left(t,\tau\right)$ is the fraction of electrons emitted up to a time $t$ and $\tau$ (determined experimentally\cite{main_Ivan}) is a function of incident photon energy.
The fraction of electrons reaching the surface with energy between $E$ and $E+dE$ and between time $t$ and $t+dt$ is
\begin{equation}
F\left(E,t,\hbar\nu\right) dE dt=f\left(E,\hbar\nu,t\right)\eta\left(t,\hbar\nu\right) dE  dt\label{eq:dist1}
\end{equation}
Consider electrons with energy $E$ in the GaAs crystal just beneath the surface. We assume that these are distributed uniformly on a sphere in $k$-space. Following the derivation in\cite{narrow_cone} the angular distribution of emitted electrons is given by
\begin{equation}
n(E,\theta)d\theta=\frac{(E+E_{A})\cos\theta d\theta}{\sqrt{2E}\left(E-\frac{m_{e}}{m_{\Gamma}}(E+E_{A})\sin^{2}\theta\right)^{1/2}}\label{eq:dist}
\end{equation}
where $E_{A}$ is the negative electron affinity and $\theta$ is the angle with respect to the local surface normal.
As a part of the MC simulation, electrons were launched in these conical distributions from the $256\times256$ grid on the surfaces shown in Fig.1. The MTE, $T(E)$ was calculated numerically as a function of the electron energy $E$ just before emission from the surface.
Finally, MTE as a function of the incident photon energy is obtained by integrating over all the energy distributions given by eq.\ref{eq:dist1}. The MTE as a function of $\hbar\nu$ is given by the integral
\begin{equation}
T(\hbar\nu)=\frac{\int\limits_{0}^{\infty}\int\limits_{0}^{\infty}T(E)F\left(E,t,\hbar\nu\right) dE dt}{\int\limits_{0}^{\infty}\int\limits_{0}^{\infty}F\left(E,t,\hbar\nu\right) dE dt}\label{eq:final}
\end{equation}
MTE as a function of the incident laser wavelength as obtained from eq.\ref{eq:final} for the rough surface is shown in Fig.\ref{fig:3a} (solid line), and well explains the experimental data.

\begin{figure}
\includegraphics[width=\linewidth,bb=50bp 185bp 655bp 562bp]{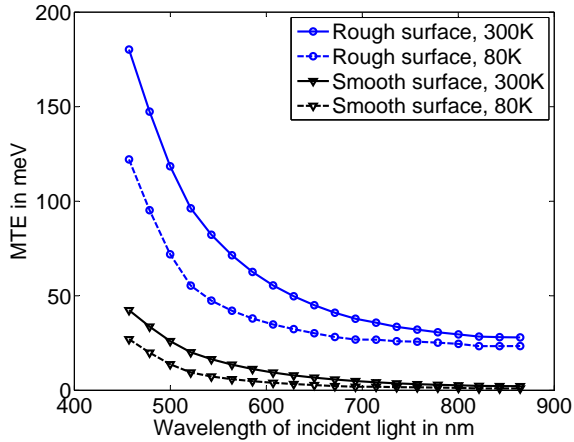}
\caption{\label{fig:2b}MTE vs wavelength of incident photons for smooth and rough surfaces at different temperatures}
\end{figure}
Fig.\ref{fig:2b} shows MTE as a function of incident wavelength calculated for the smooth and rough surfaces (Fig.1) at different temperatures. We assume that the temperature dependence comes in only from the scattering rates of the phonons. It can be seen that, MTE approaches 25meV at longer wavelenghts for the rough surface and is less than 2meV for the smooth surface. Thus, the discrepancies in the measurements of MTE can be explained by the nano-scale surface roughness due to the variations in the preparation of the bulk GaAs. Hence the surface roughness must be duly characterized to a scale of 1nm. Our results also predict a drop in MTE which is much smaller than the decrease in thermal energy upon temperature reduction. This is consistent with experimental observations\cite{transverse_spread}. This leads us to conclude that the thermal energy of electrons inside GaAs does not get directly translated into the MTE.

In summary,  a dramatic improvement in the photocathode performance for bright electron generation\cite{max_brightness} is anticipated with the proper control of the surface preparation procedures in III-V NEA photoemitters.

This work is supported by NSF DMR-0807731 and DOE DE-SC0003965.

\end{document}